\documentstyle[prd,aps,twocolumn]{revtex}
\begin{document}
\draft

\twocolumn[\hsize\textwidth\columnwidth\hsize\csname @twocolumnfalse\endcsname
\title{Quantum Nucleardynamics as an $SU(2)_N \times U(1)_Z$ Gauge Theory}
\author{Heui-Seol Roh\thanks{e-mail: hroh@nature.skku.ac.kr}}
\address{BK21 Physics Research Division, Department of Physics, Sung Kyun Kwan University, Suwon 440-746, Republic of Korea}
\date{\today}
\maketitle

\begin{abstract}
It is illustrated that quantum nucleardynamics (QND) as an $SU(2)_N \times U(1)_Z$ gauge theory,
which is generated from quantum chromodynamics (QCD) as an $SU(3)_C$ gauge theory
through dynamical spontaneous symmetry breaking, successfully describes nuclear phenomena at low energies.
The proton and neutron assigned as a strong isospin doublet are identified as a colorspin plus weak isospin doublet.
Massive gluon mediates strong interactions with the effective coupling constant
$G_R/\sqrt{2}= g_n^2/8 M_G^2 \approx 10 \ \textup{GeV}^{-2}$
just like Fermi weak constant $G_F/\sqrt{2} = g_w^2/8
M_W^2 \approx10^{-5} \ \textup{GeV}^{-2}$ in the Glashow-Weinberg-Salam model
where $g_n$ and $g_w$ are the coupling constants and $M_G$ and $M_W$ are the gauge boson masses. Several explicit evidences such as
cross sections, lifetimes, nucleon-nucleon scattering, magnetic
dipole moment, nuclear potential, gamma decay, etc. are shown in
support of QND.  The baryon number conservation is the consequence of the
$U(1)_Z$ gauge theory and the proton number conservation is the
consequence of the $U(1)_f$ gauge theory.
\end{abstract}

 ] \narrowtext

\section{Introduction}

One of the longstanding problems in physics is the establishment of nuclear dynamics
at low energies as strong interactions based on a fundamental theory. The knowledge of
the nuclear force is phenomenologically very good for $r > 2.0$ fm but it is known
only partly for $0.8 < r < 2.0$ fm and only poorly for $r < 0.8$ fm. This implies that
there is no systematic approach based on a fundamental theory applicable in the wide
range from the short distance to the long distance for nuclear physics. Quantum
chromodynamics (QCD) as an $SU(3)_C$ gauge theory \cite{Frit} is generally believed as
the fundamental theory for strong interactions and is successful in various
perturbative tests. Nevertheless, there are, in the one hand, two distinct problems in
QCD with quarks and gluons as fundamental constituents. One is the confinement, which
is not rigorously explained in the low energy region and the other is the $\Theta$
vacuum \cite{Hoof2}, which is a superposition of the various false vacua, violating CP
symmetry. At lower energies, on the other hand, many nuclear effective models as the
alternatives of QCD were proposed but their applications are not complete and limited
to a few aspects. It is thus the motivation of quantum nucleardynamics (QND), which is
derived from QCD as the consequence of the confinement and $\Theta$ vacuum, to explain
diverse nuclear phenomena at lower energies consistently. For examples, nuclear issues
to be clarified are as follows: Lande's spin g-factor for nucleon, constant nucleon
density, intrinsic quantum number, nucleon-nucleon scattering data, baryon number
conservation, proton number conservation, baryon asymmetry, etc. These phenomena at
relatively lower energies may be partly explained by effective models but each
effective model is only applicable to limited issues. To resolve these problems all
together, a new concept of colorspin is introduced so that QND as an $SU(2)_N \times
U(1)_Z$ gauge theory is dynamically spontaneous symetry broken from QCD as an
$SU(3)_C$ gauge theory in terms of the local gauge symmetry of colorspin \cite{Roh3}.
This paper attempts to demonstrate that QND is an $SU(2)_N \times U(1)_Z$ gauge theory
for nuclear interactions just as the Glashow-Weinberg-Salam (GWS) model \cite{Glas} is
an $SU(2)_L \times U(1)_Y$ gauge theory for weak interactions.

QND is generated from QCD through the dynamical spontaneous symmetry breaking (DSSB)
mechanism: $SU(3)_C \rightarrow SU(2)_N \times U(1)_Z \rightarrow U(1)_f$. QND as an
$SU(2)_N \times U(1)_Z$ gauge theory for strong interactions of nucleons and the GWS
model \cite{Glas} as an $SU(2)_L \times U(1)_Y$ gauge theory for weak interactions of
quarks and leptons have analogous properties \cite{Roh3}. The effective strong
coupling constant $G_R/\sqrt{2}= g_n^2/8 M_G^2 \approx 10 \ \textup{GeV}^{-2}$ like
Fermi weak constant $G_F/\sqrt{2} = g_w^2/8 M_W^2 \approx10^{-5} \ \textup{GeV}^{-2}$
and the color mixing angle $\sin^2 \theta_R = 1/4$ like the Weinberg mixing angle
$\sin^2 \theta_W = 1/4$ thus play important roles in nuclear interactions. The proton
and neutron assigned as a strong isospin doublet are identified as a colorspin plus
weak isospin doublet. The magnetic dipole moment and nucleon-nucleon scattering give
the explicit evidence for the color intrinsic angular momentum of hadrons. Several
explicit evidences such as cross sections, lifetimes, nucleon-nucleon scattering,
meson-nucleon scattering, magnetic dipole moment, nuclear potential, nuclear binding
energy, gamma decay, baryon asymmetry, etc. are shown in support of QND. The nuclear
radius $r = r_0 A^{1/3} = r_0 n^2$ with the nucleon number $A$ introduces the
extrinsic principal quantum number $n = A^{1/6}$ and the extrinsic angular momentum
$l$ originated from color charges. It is suggested that the baryon number conservation
is the consequence of the $U(1)_Z$ gauge theory and the proton number conservation is
the consequence of the $U(1)_f$ gauge theory.

This paper is organized as follows. In Section II,
QND as an $SU(2)_N \times U(1)_Z$ gauge theory with the $\Theta$ term
is introduced as the consequence of DSSB of $SU(3)_C$ gauge theory.
The conclusive evidences and applications of QND in various nuclear phenomena are discussed
in Section III. Section IV describes comparison between QND and effective
models.  Section V is devoted to conclusions.

\section{Dynamical Spontaneous Symmetry Breaking in Strong Interactions}

The DSSB mechanism of QCD to QND is briefly described before discussing the
evidences of QND in nuclear dynamics:
According to the reference \cite{Roh3}, $SU(3)_C \rightarrow SU(2)_N \times U(1)_Z \rightarrow U(1)_f$.

The $SU(3)_C$ gauge-invariant Lagrangian density with the $\Theta$ term is, in four vector notation, given by
\begin{equation}
\label{qchr} {\cal L}_{QCD} = - \frac{1}{2} Tr  G_{\mu \nu} G^{\mu \nu} + \sum_{i=1}
\bar \psi_i i \gamma^\mu D_\mu \psi_i + \bar \Theta \frac{g_s^2}{16 \pi^2} Tr G^{\mu
\nu} \tilde G_{\mu \nu}
\end{equation}
where the subscript $i$ stands for the classes of pointlike spinors, $\psi$ for the
spinor and $D_\mu = \partial_\mu - i g_s A_\mu$ for the covariant derivative with the
coupling constant $g_s$. Particles carry the local $SU(3)_C$ charges and the gauge
fields are denoted by $A_{\mu} = \sum_{a=0}^8 A^a_{\mu} \lambda^a /2$ with the
Gell-Mann matrices $\lambda^a, a=0, . . , 8$. The field strength tensor is given by
$G_{\mu \nu} = \partial_\mu A_\nu - \partial_\nu A_\mu - i g_s [A_\mu, A_\nu]$ and
$\tilde G_{\mu \nu}$ is the dual field strength tensor. In the Lagrangian density, the
explicit quark mass term is not contained but the $\Theta$ vacuum term is added. The
fine structure constant
\begin{math}
\label{fins}
\alpha_s = g_s^2 / 4 \pi
\end{math}
is the only free parameter.

Gluon interactions in the effective $SU(3)_C$ gauge invariant Lagrangian density with the bare $\Theta$ term are
\begin{equation}
\triangle {\cal L}^e = - \frac{1}{2} Tr  G_{\mu \nu} G^{\mu \nu} + \Theta
\frac{g_s^2}{16 \pi^2} Tr G^{\mu \nu} \tilde G_{\mu \nu} .
\end{equation}
Apart from charge nonsinglet gauge bosons, four singlet gauge boson interactions are
parameterized by the $SU(3)$ symmetric scalar potential:
\begin{equation}
\label{higs}
V_e (\phi) = V_0 + \mu^2 \phi^2 + \lambda \phi^4
\end{equation}
which is the typical potential with $\mu^2 < 0$ and $\lambda > 0$
for spontaneous symmetry breaking. The first term of the right
hand side corresponds to the vacuum energy density representing
the zero-point energy by non-axial singlet bosons. The axial
vacuum field $\phi$ is shifted by an invariant quantity $\langle
\phi \rangle$, which satisfies
\begin{math}
\langle \phi \rangle^2 = \phi_0^2 + \phi_1^2 + \phi_2^2 + \phi_3^2
\end{math}
with the condensation of the axial singlet gauge boson $\langle \phi \rangle = (\frac{- \mu^2}{2
\lambda})^{1/2}$. DSSB is connected with the surface term $\Theta \frac{g_s^2}{16
\pi^2} Tr G^{\mu \nu} \tilde G_{\mu \nu}$, which explicitly breaks down the $SU(3)_C$
gauge symmetry through the condensation of axial singlet gauge bosons and breaks
down the axial current. The $\Theta$ can be assigned by an dynamic parameter by
\begin{equation}
\label{thev}
\Theta = 10^{-61} \ \rho_G /\rho_m
\end{equation}
with the matter energy density $\rho_m$ and the vacuum energy density
$\rho_G = M_G^4$.

The Lagrangian density of QND as an $SU(2)_N \times U(1)_Z$ gauge theory has the same
form with QCD as an $SU(3)_C$ gauge theory without the explicit mass term:
\begin{equation}
{\cal L}_{QND} = - \frac{1}{2} Tr  G_{\mu \nu} G^{\mu \nu}
+ \sum_{i=1}  \bar \psi_i i \gamma^\mu D_\mu \psi_i  + \Theta \frac{g_n^2}{16 \pi^2} Tr G^{\mu \nu} \tilde G_{\mu \nu},
\end{equation}
where the bare $\Theta$ term \cite{Hoof2} is a nonperturbative term added to the
perturbative Lagrangian density with an $SU(2)_N \times U(1)_Z$ gauge invariance. The
$\Theta$ term apparently odd under both P, T, C, and CP operation. The nuclear
coupling constant $g_n^2 = c_f g_s^2 = \sin^2 \theta_R g_s^2 = g_s^2/4$ is given in
terms of the strong coupling constant $g_s$ and the color factor $c_f$. The effective
strong coupling constant at low energy is expressed in analogy with the
phenomenological, electroweak coupling constant $G_F$:
\begin{equation}
\frac{G_R}{\sqrt{2}} = - \frac{c^n_f g_s^2}{8 (k^2 - M_G^2)} \simeq \frac{c^n_f g_s^2}{8 M_G^2}
\end{equation}
where $M_G$ indicates the mass of gluon, $k$ denotes the four momentum, and $c^n_f =
\sin^2 \theta_R$ represents the nuclear color factor. The gluon mass is thus generally
reduced by the singlet gluon condensation $\langle \phi \rangle$:
\begin{equation}
\label{glma} M_G^2 = M_H^2 - c_f g_s^2 \langle \phi \rangle^2 = c_f g_s^2
[A_{0}^2 - \langle \phi \rangle^2]
\end{equation}
where $M_H = \sqrt{c_f} g_s A_{0}$ is the gauge boson mass at the grand unification
scale, $A_{0}$ is the singlet gauge boson, and $\langle \phi \rangle$ represents the condensation of
the axial singlet gauge boson. The color factor $c_f$ becomes the
symmetric factor with even parity for singlet gauge boson and is the asymmetric factor
with odd parity for axial singlet gauge boson. The vacuum energy due to the zero-point
energy, represented by the gauge boson mass, is thus reduced by the decrease of the
color factor and the increase of the axial singlet gluon condensation as temperature
decreases. This process makes the breaking of discrete symmetries P, C, T, and CP due to the gluon mass.

\section{Quantum Nucleardynamics as a Gauge Theory}

In QND, the most important concepts are that nucleons are treated as a color doublet
and gluons are massive as hinted by the analogy of the GWS model \cite{Glas}.

QCD as the color $SU(3)_C$ symmetry generates quantum nucleardynamics (QND) as
the $SU(2)_N \times U(1)_Z$
symmetry, which governs nuclear strong dynamics, and the $U(1)_f$
symmetry, which governs nuclear electromagnetic dynamics, with the
condensation of singlet gluons. The $SU(2)_N \times U(1)_Z$
symmetry and $U(1)_f$ symmetry using the symmetric color factors,
$c^s_f = (c_f^b, c_f^n, c_f^z, c_f^f) = (1/3, 1/4, 1/12, 1/16)$,
are applied to the typical strong interactions: color asymmetric
configuration with odd parity is not observed but color symmetric
configuration with even parity is observed. The factors described
above are the pure color factors due to color charges but the
effective color factors used in nuclear dynamics must be
multiplied by the isospin factor $i_f^w = \sin^2 \theta_W = 1/4$
with the weak Weinberg angle $\theta_W$ since the proton and
neutron are an isospin doublet as well as a color doublet:
$c_f^{eff} = i_f^w c_f = i_f^w (c_f^b, c_f^n, c_f^z, c_f^f) =
(1/12, 1/16, 1/48, 1/64)$ for symmetric configurations. Note that
the color mixing is the origin of the Cabbibo angle for quark
flavor mixing in weak interactions. For example, the
electromagnetic color factor for the $U(1)_f$ gauge theory becomes
$\alpha_f^{eff} = \alpha_s/64 \simeq 1/137$ when $\alpha_s \simeq
0.48$ at the strong scale \cite{Hinc}. Nucleons as spinors possess up and down
colorspins as a color doublet just like up and down strong
isospins:
\begin{equation}
{\uparrow \choose \downarrow}_c, \ \uparrow = {1 \choose 0}_c, \ \downarrow = {0
\choose 1}_c .
\end{equation}
This implies that conventional, global $SU(2)$ strong isospin
symmetry \cite{Heis} is postulated as the combination of local
$SU(2)$ colorspin and local $SU(2)$ weak isospin symmetries: the
static $SU(3)$ quark model is extended as the combination of
$SU(2)_N \times U(1)_Z$ gauge theory for nuclear interactions and
$SU(2)_L \times U(1)_Y$ gauge theory for weak interactions \cite{Glas}. Color
charge quantum numbers for the proton and neutron, which will be
discussed more, are shown in Table \ref{coqu} where the subscript
$d$ denotes the colorspin doublet and the subscript $s$ denotes
the colorspin singlet. Nucleons as the color spin doublet are
governed by QND as the $SU(2)_N \times U(1)_Z$ gauge theory just
as leptons or quarks as the isospin doublet are governed by the
GWS model as the $SU(2)_L \times U(1)_Y$ gauge theory in weak
interactions. This is also compatible with the confinement of the
color electric monopole inside the hadron space but the
confinement of the color magnetic monopole inside the vacuum
space. The overall wave function for a nucleon may thus be
expressed by $\psi_N = \psi (\textup{colorspin}) \psi
(\textup{isospin}) \psi (\textup{spin}) \psi (\textup{space})$.

There are several longstanding problems in nuclear physics, which
may be resolved if QND as an $SU(2)_N \times U(1)_Z$ gauge theory
is applied. The typical evidences discussed are multipole expansion,
lifetime and cross section, excitation and decay, conservation and
violation of symmetries, strong isospin and colorspin, nuclear
mass and charge, nuclear magnetic dipole moment, deuteron,
nucleon-nucleon scattering, meson-nucleon scattering, strong
isospin and colorspin, nuclear potential, shell model, and nucleus
binding energy.

\subsection{Intrinsic and Extrinsic Multipole Expansion}

There are two kinds of quantum numbers, intrinsic and extrinsic
quantum numbers, described in the reference \cite{Roh3}. Intrinsic
quantum numbers mainly play important role inside the hadron size
while extrinsic quantum numbers play important role outside the
hadron size. There are, on the other hand, two kinds of strong
interactions; one produces the Coulomb potential for effectively
massless gluons and the other one produces the Yukawa potential
for massive gluon. The Coulomb potential is a special case of the
Yukawa potential with the zero gauge boson mass. The Yukawa
potential derives the confinement of quarks and gluons in the
region of low energies. The effective interaction amplitude
becomes ${\cal M} = \sqrt{2} G_R J^\mu J_\mu^\dagger$ where only
the color vector current is conserved. The effective coupling
constant of the Yukawa potential, $G_R = \sqrt{2} c_f g_s^2/8
M_G^2$, is not a fixed value but it changes according to the
condensation of singlet gauge bosons; as energy goes down it
increases. The gluon mass is the QCD cutoff energy at QCD
confinement phase transition. Therefore, the interaction range is
roughly the size of the hadron $1$ fm; the interaction range by
the massive gluon is the QCD cutoff scale. According to the hadron
size, the gluon mass becomes a value around $300$ MeV for the
$SU(2)_N$ gauge theory or $140$ MeV for the $U(1)_Z$ gauge theory
in hadron formation at the QCD cutoff scale.

Inside the hadron, the Yukawa potential is decomposed with several terms corresponding multipoles:
for a few terms of lower orders, the constant potential ($\sim \Lambda_{QCD}$), the linearly increasing potential ($\sim \alpha r = \Lambda_{QCD}^2 r/2$),
the harmonic oscillator potential ($\sim \beta r^2/2 = \Lambda_{QCD}^3 r^2/6$), etc.
These terms are dependent on intrinsic color angular momenta.
The constant energy as the monopole potential due to the gluon mass is relevant for the vacuum energy:
the vacuum energy density $M_G^4 = \Lambda_{QCD}^4$ corresponds to the bag constant in the bag model.
The linearly increasing potential as the dipole potential is the dominant contribution to quark and gluon confinement:
the coefficient is $\alpha \simeq \Lambda_{QCD}^2/2 \approx 0.05 \
\textup{GeV}^2$ with $\Lambda \simeq 0.3$ GeV.
The harmonic oscillator potential as the quadrupole potential, which has the coefficient $\beta = \Lambda_{QCD}^3 /3 \approx 0.01 \ \textup{GeV}^3$, leads
to the dynamics for photons as quanta.

Outside the hadron, the Yukawa potential is also decomposed with several terms corresponding multipoles:
for a few terms of low orders, the Coulomb potential ($\propto - 1/r$), the dipole potential ($\propto 1/r^2$), the quadrupole potential ($\propto - 1/r^3$), etc.
These terms are dependent on extrinsic angular momenta.
The Coulomb potential as the monopole potential represents the repulsive interaction for proton charges with the energy $E \leq \Lambda_{QCD}$.
This term is the origin of the electromagnetic potential in a nucleus.
Dipole and quadrupole potentials deform a nucleus from the spherical shape.
Due to the requirement of even parity for nucleons, the monopole moment and quadrupole
moment are left as electric moments while the dipole moment and octopole moment
are left as magnetic moments.

Nuclear strong interactions are governed by an $SU(2)_N \times U(1)_Z$ gauge theory due to massive gluons
while nuclear electromagnetic interactions are governed by a $U(1)_f$ gauge theory due to massless photons.
Massive gluons in the vacuum space is spatially quantized by the maximum wavevector mode
$N_R \approx 10^{30}$, the total gluon number $N_G = 4 \pi N_R^3/3 \approx 10^{91}$,
and the gluon number density
$n_G = \Lambda_{QCD}^3 \approx 10^{-2} \ \textup{GeV}^3 \approx 10^{39} \ \textup{cm}^{-3}$ \cite{Roh3}.
Baryon matter represented by massive baryons is quantized by the maximum wavevector mode (Fermi mode)
$N_F \approx 10^{26}$ and the total baryon number $B = N_B = 4 \pi N_F^3/3 \approx 10^{78}$.
Massless photons are quantized by the maximum wavevector mode
$N_\gamma \approx 10^{29}$ and the total photon number $N_{t \gamma} = 4 \pi N_\gamma^3/3 \approx 10^{88}$.
Massless phonons in the matter space are spatially quantized by the maximum wavevector mode (Debye mode)
$N_D \approx 10^{25}$ and the total phonon number $N_{t p} = 4 \pi N_D^3/3 \approx 10^{75}$.
The QCD confinement due to massive gluons may be interpreted as the dual Meissner effect;
the color electric field is confined and interacts pointlikely at the longer range than the hadron size.
Nuclear force thus interacts in terms of the interchange of the massive gluon with color charges.
This implies that nuclear force interacts via the Yukawa potential due to the massive gluon both inside and outside the hadron size.

\subsection{Nuclear Lifetime and Cross Section as an $SU(2)_N \times U(1)_Z$ Gauge Theory}

There are several, explicit examples about lifetimes and cross
sections, supporting QND as an $SU(2)_N \times U(1)_Z$ gauge
theory in analogy with the GWS model as an $SU(2)_L \times U(1)_Y$
gauge theory.

Quark-antiquark pair annihilation $q + \bar q \rightarrow g + g$ gives the decay rate \cite{Grif}
\begin{equation}
\Gamma = \frac {8 \pi}{3} (\frac{g_s^2}{m_q})^2 |\psi (0)|^2
\end{equation}
in the spin, color singlet configuration with the quark at rest.
Using the values $\alpha_s = 0.48, \ m_q = 310$ MeV, and $|\psi (0)|^2 = (148 \ \textup{MeV})^3$, the decay rate $\Gamma = 135$ MeV or
the lifetime $\tau = 3.6 \times 10^{-24}$ s is obtained.
The state $\Sigma^0$ is formed as a resonance of central mass $1385$ MeV in a $K^{-} p$ interaction:
\begin{equation}
K^- + p \rightarrow \Sigma^0 \rightarrow \Lambda + \pi^0
\end{equation}
where the Q-value in the decay $130$ MeV and the lifetime $\tau = 1/\Gamma \approx 10^{-23}$ s are estimated from the measured
decay width $\Gamma = 36$ MeV.
If the decay rate $\Gamma \simeq G_R^2 m_\Sigma^5$ is used in analogy with the muon decay rate
$\Gamma = G_F^2 m_\mu^5/192 \pi^3$ of weak interactions, the
$\Sigma^0$-hyperon lifetime becomes the order of $\tau \approx 10^{-23}$ s
compared with the muon lifetime in the order of $10^{-6}$ s.
In the strong decay process, the exchange of the massive gluon is taken into account
like the exchange of massive intermediate vector boson in the weak decay process.
The strong decay $\Delta \rightarrow n + p$ and the weak decay
$\Sigma^{\pm} \rightarrow n + \pi^{\pm}$ have almost same $0.12$ GeV kinetic energy, approximately.
Their lifetime ratio becomes
\begin{equation}
\frac{\tau(\Delta \rightarrow n + p)}{\tau(\Sigma \rightarrow n + \pi)}
\simeq \frac{1/G_R^2 m_\Delta^5}{1/G_F^2 m_\Sigma^5}
\simeq \frac{10^{-23} \ \textup{s}}{10^{-10} \ \textup{s}} .
\end{equation}
Similarly, the lifetime ratio between
$\Sigma^0 \rightarrow \Lambda + \pi^0$ and $\Sigma^- \rightarrow n + \pi^-$
is found to be
\begin{equation}
\frac{\tau(\Sigma^0 \rightarrow \Lambda + \pi^0)}{\tau(\Sigma^- \rightarrow n + \pi^-)}
\simeq \frac{1/G_R^2 m_{\Sigma^0}^5}{1/G_F^2 m_{\Sigma^-}^5}
\simeq \frac{10^{-23} \ \textup{s}}{10^{-10} \ \textup{s}} .
\end{equation}
Since the typical cross section in weak interactions can be estimated by $\sigma \simeq G_F^2 T^2 \approx 10^{-44} \ \textup{m}^2$,
the typical cross section in strong interactions is $\sigma \simeq G_R^2 T^2 \approx 10^{-30} \ \textup{m}^2$, which gives good agreement with experiment result.
In the decay of $\Delta^{++}$, which hints the color quantum number,
$\Delta^{++} \rightarrow \pi^+ + p$, the lifetime of $\Delta^{++}$ is $10^{-23}$ s.
This can be interpreted by the distance of about $1$ fm estimated by the
gluon mass of about $300$ MeV in this scheme.
In the strong interaction of $\pi + p \rightarrow \pi + p$ and the weak
interaction of $\nu + p \rightarrow \nu + p$, the cross section ratio
around the kinetic energy $T = 1$ GeV becomes
\begin{equation}
\frac{\sigma(\pi + p \rightarrow \pi + p)}{\sigma(\nu + p \rightarrow \nu + p)}
\simeq \frac{G_R^2 T^2}{G_F^2 T^2}
\simeq \frac{10 \ \textup{mb}}{10^{-11} \ \textup{mb}} .
\end{equation}
These examples described above explicitly show the typical lifetimes and cross sections in strong interactions as expected.

\subsection{Nuclear Excitation and Decay as a $U(1)_f$ Gauge Theory}

QND as a $U(1)_f$ gauge theory, where the photon is the massless gauge boson
and the proton is positively charged, is proposed just like QED as a $U(1)_e$ gauge theory,
where the photon is the massless gauge boson and the electron is negatively
charged, is used.
Electromagnetic dynamics in nucleus as a $U(1)_f$ gauge theory is applied to evaluate nuclear excitation and decay
around the order of $1$ MeV.

The coupling constant $\alpha_f = \alpha_s/16$ for a $U(1)_f$ gauge theory at the
strong scale is used to evaluate excitation levels. The pure color coupling constant
$\alpha_f = \alpha_s/16$ has about four times stronger than the coupling constant
$\alpha_e = \alpha_i/16$ for a $U(1)_e$ gauge theory at the weak scale: the effective
coupling constant $\alpha_f^{eff} = i^w_f \alpha_f = \alpha_s/64 = \alpha_e$. The
emission of energetic photons as gamma radiation is typical for a nucleus deexcitation
from some high lying excited state to the ground state. This represents a reordering
of the nucleon in the nucleus with a lowering of mass from the excited mass to the
lowest mass. Electromagnetic transition due to the proton charge and gamma decay are
well established subjects. Alpha decay is a Coulomb repulsion effect, which is
important for heavy nuclei since the disruptive Coulomb force increases with size at a
faster rate than the nuclear binding force. For another example, nuclei with closed
shell plus one valence nucleon is considered in analogy with the hydrogen atom. The
Coulomb potential originated from color charges is thus automatically realized in
nucleon-nucleon interactions in terms of the $U(1)_f$ gauge theory in addition to the
Coulomb potential originated from isospin electric charges. The separation energy of a
valence nucleon is $m_n \alpha_f^2/2 \simeq 0.42$ MeV with the coupling constant
$\alpha_f \simeq 0.03$: $m_n \alpha_f^2/2 n^2$ is the energy level with the principal
quantum number $n$. The colorspin-orbit coupling energy of nuclei has the order of
$m_n \alpha_f^4 \simeq 0.8$ KeV and the nuclear Lamb shift between states with the
same total but the different angular momentum has the energy difference of about $m_n
\alpha_f^5 \simeq 0.02$ KeV. The Lamb shift in nucleus may be evaluated if the
radiative correction due to the loop contribution as a $U(1)_f$ gauge theory is
included.

\subsection{Conservation and Violation in Nuclear Dynamics}

The proton number conservation is the consequence of the $U(1)_f$ gauge theory just as
the electron number conservation is the consequence of the $U(1)_e$ gauge theory and
the baryon number conservation is the result of the $U(1)_Z$ gauge theory for strong interactions just as
the lepton number conservation is the result of the $U(1)_Y$ gauge theory for weak interactions.
Discrete symmetries are perturbatively conserved but are nonperturbatively violated in strong interactions.

\subsubsection{Conservation of Proton Number and Baryon Number}

The conservation of the proton number is the result of the
$U(1)_f$ local gauge theory just as the conservation of the
electron number is the result of the $U(1)_e$ local gauge theory.
The charge quantization is given by $\hat Q_f = \hat C_{3} + \hat
Z_c/2$ where $\hat C_3$ is the third component of the colorspin
operator $\hat C$ and $\hat Z_c$ is the hyper-color charge
operator. This form has the analogy with the electric charge
quantization $\hat Q_e = \hat I^w_{3} + \hat Y^w/2$ \cite{Glas} in
weak interactions and $\hat Q_e = \hat I^s_{3} + \hat Y^s/2$
\cite{Gell} in the quark model. The hyper-color charge operator
may be defined by $\hat Z_c = \hat B + \hat S + \hat C + \cdot
\cdot$ with the baryon number operator $\hat B$, the strangeness
number operator $\hat S$, and the charm number operator $\hat C$
just as the hypercharge operator is defined by $\hat Y = \hat B -
\hat L$ with the baryon number operator $\hat B$ and the lepton
number operator $\hat L$. Color charge quantum numbers for the
proton and neutron are shown in Table \ref{coqu} where nucleons
are classified by color doublet and color singlet. Colorspin
doublet nucleons are governed by the $SU(2)_N \times U(1)_Z$ gauge
theory just as isospin doublet leptons or quarks are governed by
the $SU(2)_L \times U(1)_Y$ gauge theory in weak interactions. As
noted by the observation of the proton decay, the lifetime of the
proton is more than $10^{32}$ years. The conservation of the
baryon number is analogous to the conservation of the proton
number. The conservation of the baryon number is the consequence
of the $U(1)_Z$ local gauge theory just as the conservation of the
lepton number is the consequence of the $U(1)_Y$ local gauge
theory. Table \ref{coga} shows relations between conservation laws
and gauge theories. Baryons are conserved as the colorspin doublet
but are not conserved as the color singlet; this is analogous to
the conservation of leptons as the isospin doublet but the
nonconservation of leptons as the isospin singlet in weak
interactions. The immediate result of the proton number
conservation or the baryon number conservation is shown in the
mass density and charge density of nuclear matter. The effective
charge unit of the charge operator $\hat Q_f$ is $q^{eff}_f =
\sqrt{\pi \alpha_s/64}$ while the charge unit of the charge
operator $\hat Q_e$ is $e = \sqrt{\pi \alpha_i/4}$: the absolute
magnitude of $q^{eff}_f$ is the same with that of $e$ since
$\alpha_s \simeq 0.48$ at the strong scale and $\alpha_i \simeq
0.12$ at the weak scale. The concept of the effective charge
$q_f^{eff}$ is consistent with one of colorspin and isospin
intrinsic angular momenta. Note that there is no discrete symmetry
breaking due to the $U(1)_f$ local gauge theory.

Conservation laws of the proton and baryon can be applied in analyzing nuclear
interactions including nuclear scattering and reaction as well as conservation laws of
the total energy and linear momentum. Angular momentum conservation law enables
interacting particles to assign the intrinsic angular momenta.

\subsubsection{Violation of Discrete Symmetries}

Discrete symmetries are perturbatively conserved in strong
interactions. The violation of discrete symmetries is thus the
nonperturbative indication of DSSB from an $SU(3)_C$ gauge theory
to an $SU(2)_N \times U(1)_z$ gauge theory and then to a $U(1)_f$
gauge theory. P violation is manifest since there are no
parity partners of pseudoscalar mesons, vector mesons, and
baryons. This resolves the $U(1)_A$ problem for the
non-observation of the sigma meson as the parity partner of the
pion. The fact that the color doublet vector current of the proton
and neutron is conserved but the color singlet axial current is
not conserved indicates parity violation. The evidence of CP and T
violation appears in the magnitude of the $\Theta$ constant, which
is measured by the electric dipole moment of the neutron: $\Theta
< 10^{-9}$ \cite{Alta}. C violation predicts that the number ratio of
antibaryons to baryons is extremely small in the matter space:
this is connected with the baryon-antibaryon number asymmetry
$\delta_B \simeq 10^{-10}$ \cite{Stei0}.

\subsection{Strong Isospin and Colorspin}

The proton and neutron are considered as two different
manifestations of the nucleon in terms of strong isospin symmetry
introduced by Heisenberg \cite{Heis}. Strong isospin has formal
analogy with ordinary spin but it has nothing to do with rotation
with the coordinate space, contrary to spin. However, if
approximate strong isospin symmetry is replace with the mixed
symmetry of colorspin and weak isospin, a nucleon state possesses
local colorspin, weak isospin, and spin degrees of freedom in
intrinsic coordinate space since colorspin and weak isospin are
intrinsic angular momenta like spin angular momentum. The effects
of colorspin will thus appear in the magnetic dipole moment and
electric quadrupole moment in nucleons and they may be
investigated by QND. Quantum numbers of general nucleon-nucleon
systems are summarized in Table \ref{nnqu} when colorspin degrees
of freedom are taken into account in addition to isospin and spin
degrees of freedom; there exist six distinctive states because of
two colorspins, two weak isospins, and two spins.

Quark model, isospin quantization, colorspin quantization, and colorspin
applications are addressed in the following.

\subsubsection{Quark Model, Isospin, and Colorspin}

The quark model introduced by Gell-Mann and Zweig is extremely successful in describing static hadrons.
In this part, its relation to the GWS model \cite{Glas} and QND is considered.
As described above, quarks possess three types of intrinsic charges, color, isospin, and spin.
Color charge and weak isospin charge are related to strong isospin charge.
The quark model with strong isospins as charges is applied to static hadrons composed of quarks,
which are governed by the GWS model with weak isospin charges
at the weak energy scale and QND with color charges at strong scale.
QND and GWS model are the dynamical extensions of the static quark model as a local gauge theories.
$SU(2)$ strong isospin is thus regarded as just the combination of $SU(2)$ colorspin and $SU(2)$ weak isospin:
for example, the proton with strong isospin $1/2$ consists of three quarks (uud) and the neutron with strong isospin $-1/2$ consists of
three quarks (udd) where u has colorspin and weak isospin and d has colorspin and weak isospin simultaneously.
The charge quantization $\hat Q_e = \hat I^s_3 + \hat Y^s/2$ in the quark model is
analogous to the charge quantization $\hat Q_e = \hat I^w_3 + \hat Y^w/2$ in weak interactions where $\hat I^s$ is the strong isospin operator,
$I^w$ is the weak isospin operator, $\hat Y^s$ is the strong hypercharge quantum operator,
and $\hat Y^w$ is the weak hypercharge quantum operator.
$\hat Y^s = \hat B + \hat S + \hat C$ is decomposed of
the baryon number $B$, the strangeness number $S$, and the charm number $C$.
The longitudinal component of strong isospin is expressed by $I_3^s = (Z-A)/2$ in terms of the nuclear number $Z$ and the mass number $A$.
The color electric charge is similarly quantized by $\hat Q_f = \hat C_3 + \hat Z_c/2$ to describe the proton charge
where $C_3$ is the longitudinal component of colorspin $C$ and $Z_c$ is the hyper-color charge.
The hyper-color charge operator may be defined by $\hat Z_c = \hat B + \hat S + \hat C + \cdot \cdot$
with the baryon number $B$, the strangeness number $S$, and the charm number $C$
just as the weak hypercharge operator is defined by $\hat Y^w = \hat B - \hat L$
with the baryon number $B$ and the lepton number $L$.
Overall, the combination of the GWS model and QND as quantum flavordynamics is the dynamical extension of the static quark model for hadrons.
This might be a clue for the mass difference between constituent quarks and current quarks.
Current quarks with the mass around $5$ MeV possess the bigger difference number $N_{sd}$, presumably $64 = (2 \times 2)^3$ times bigger
from colorspin and isospin degrees of freedom,
than constituent quarks since current quarks have combined triplet color, triplet isospin, and doublet spin degrees of freedom and constituent quark has
doublet colorspin, doublet isospin, and doublet spin degrees of freedom.

\subsubsection{Applications of Colorspin}

Clear examples of the colorspin assignments can be found in
nuclear analog states with the same $A$ but different $N$ and $Z$.
For example, the nuclei with the $A=14$ system, $^{14}_{\
6}C_{8}$, $^{14}_{\ 7}N_{7}$, and $^{10}_{\ 8}O_{6}$ are
considered. Since they are conventional strong isospin triplet
states ($i^s = 1$), all three nuclei have the almost same
energies, apart from the electromagnetic energy: the respective
shifts from the strong isospin singlet state of $^{14}_{\ 7}N_{7}$
are $2.36$ MeV in $^{14}_{\ 6}C_{8}$, $2.31$ MeV in $^{14}_{\
7}N_{7}$, and $2.44$ MeV in $^{10}_{\ 8}O_{6}$ \cite{Ajze}. The
strong isospin triplet energy in $^{14}_{\ 7}N_{7}$ is slightly
different due to the mixing of two states with quantum numbers
$(i=1, s=0, c=1)$ and $(i=1, s=1, c=0)$. The strong isospin
singlet state of $^{14}_{\ 7}N_{7}$ is also the mixed state of two
states with quantum numbers $(i=0, s=1, c=0)$ and $(i=0, s=0,
c=1)$. Similar arguments are applicable to the three isobar nuclei
$^{10}_{\ 4}Be_{6}$, $^{10}_{\ 5}B_{5}$, and $^{10}_{\ 6}C_{4}$
and to the three isobar nuclei $^{6}_{2}He_{4}$, $^{6}_{3}Li_{3}$,
and $^{6}_{4}Be_{2}$. Further applications of Table \ref{nnqu}
will be given in the deuteron state and the nucleon-nucleon
scattering.

The reason for the mass difference for hadrons may come from the
contribution of colorspin or isospin charges. For example, quarks
u and d are an isospin doublet but their masses are slightly
different. The mass difference, $m_u \neq m_d$, is the isospin
violation by strong interactions, in which color degrees of
freedom play important role. It might provide the clue for the
mass difference between isospin multiplets in hadrons, which needs
to be further speculated. The mass difference of the proton (uud)and
neutron (udd)($\sim 1.3$ MeV) has three possible reasons related
to colorspin-colorspin and isospin-isospin interactions: the
difference in masses of the u and d quarks $(m_d > m_u)$, the
energy difference in electric and magnetic interactions. The mass
difference from electromagnetic interactions mainly comes from the
difference in the u-u quark interaction of the proton and the d-d
quark interaction of the neutron since u-d interactions are common
in both nucleons. Coulomb energy difference between the u-u and
d-d quark interaction becomes $[(2/3)^2 - (-1/3)^2] m_u
\alpha_e^2/2 \simeq 0.5$ MeV and magnetic energy difference
becomes $[(2/3)^2 - (-1/3)^2] 2 \pi \alpha_e |\psi (0)|^2/3 m_u^2
\simeq 0.5$ MeV where $2/3$ represents the electric charge of the
u quark and $-1/3$ does the electric charge of the d quark.
According to Dashen sum rule, $m^2(\pi^\pm) - m^2(\pi^0) =
m^2(K^\pm) - m^2(K^0)$ if the electromagnetic interaction is the
only source for the mass difference of isospin multiplets.
However, the fact that the mass difference sign between two
isospin  multiplets is wrong indicates another interaction for the
difference: $\Delta m(K^0 - K^\pm) \simeq 4$ MeV and $\Delta
m(\pi^0 - \pi^\pm) \simeq - 4.6$ MeV. Similarly, $\Delta
m(\Sigma^0 - \Sigma^+) \simeq 3.1$ MeV, $\Delta m(\Sigma^0 -
\Sigma^-) \simeq -4.9$ MeV, and $\Delta m(\Xi^0 - \Xi^\pm) \simeq
-6.5$ MeV are observed. These examples show the explicit
requirement of color interactions due to color degrees of freedom
in addition to electromagnetic interactions due to isospin degrees
of freedom. Note that the coupling constant for pure color charges
is stronger than that for pure isospin charges. For instance,
three pion states form a color triplet but the neutral pion is the
mixed state of color triplet and color singlet, which makes the
lower mass compared with the charge pions.

Another example for mixing between color and isospin degrees of freedom is the
Cabbibo angle \cite{Cabb} as the mixing angle between d quark and s quark with the same electric charge $-e/3$.
The Kobayasi-Maskawa matrix \cite{Koba} is a more extended version of the flavor mixing due
to intrinsic color charges.

\subsection{Nuclear Mass, Charge, and Size: Principal Quantum Number}

According to the multipole expansion of the Yukawa potential, there exists the constant potential as the intrinsic monopole term
depending on the gauge boson mass $M_G$, which is the origin of the almost constant mass
and there exists the Coulomb potential as the extrinsic monopole term
depending on the color factors, which is the origin of the almost constant charge.
They are related to the $U(1)_Z$ gauge theory and $U(1)_f$ gauge theory respectively.
In this approach, the QCD vacuum and baryon matter energies are spatially quantized as well as photon
and phonon energies.
These total particle numbers $N_G \simeq 10^{91}$, $N_B \simeq 10^{78}$,
$N_{t \gamma} \simeq 10^{81}$, and $N_{t p} \simeq 10^{75}$ are conserved good quantum numbers.

Nuclear matter is quantized by the maximum wavevector mode
$N_F \approx 10^{26}$ and the total baryon number $B = N_B = 4 \pi N_F^3/3  \approx 10^{78}$
as the consequence of the baryon number conservation or the baryon
asymmetry $\delta_B \simeq 10^{-10}$ \cite{Stei0}.
Baryon matter quantization is consistent with the
nuclear number density $n_n = n_B = A/(4 \pi r^3/3) \approx 1.95 \times 10^{38} \ \textup{cm}^{-3}$
with the nuclear mass number $A$ and the nuclear matter radius $r$ at the strong scale
$M_G \simeq 10^{-1}$ GeV and is consistent with
Avogadro's number $N_A = 6.02 \times 10^{23} \ \textup{mol}^{-1} \approx 10^{19} \ \textup{cm}^{-3}$
at the atomic scale $M_G \simeq 10^{-8}$ GeV:
the nuclear matter density is comparable to massive gluon density $M_G^3 \simeq 10^{38} \ \textup{cm}^{-3}$ at the strong scale
and atomic matter density is also comparable to massive gauge boson density $M_G^3 \simeq 10^{19} \ \textup{cm}^{-3}$ at the atomic scale.
This is also compatible with
the nuclear number density $n_B = 2 k_F^2/3 \pi^2$ with the Fermi momentum of a free nucleon $k_F = 1.33 \ \textup{fm}^{-1}$ and
the Fermi energy $\epsilon_F = k_F^2/2 m_n \approx 37$ MeV and compatible with the nucleus radius
\begin{equation}
\label{nura}
r = r_0 A^{1/3}
\end{equation}
with the nucleon radius $r_0 \approx 1.2$ fm.
In analogy with the electron orbit radius of the hydrogen atom $r_e = a_0 n_e^2$ with
the atomic radius $a_0 = 1/2 m_e \alpha_y$ or
the Bohr radius $a_B = 1/m_e \alpha_e = 5.29 \times 10^{-9}$ cm and the principal quantum number $n_e$,
the above equation (\ref{nura}) may be written by
\begin{equation}
\label{nura1}
r = r_0 n^2
\end{equation}
where the nucleon mass radius is $r_0 = 1/2 m_n \alpha_z \approx 1.2$ fm with the color degeneracy factor $2$
and the principal quantum number is $n = A^{1/6} = B^{1/6}$ with the baryon number $B$.
The baryon matter density $\rho_B$ may be connected with a baryon coupling constant $\alpha_z$ by
\begin{equation}
\rho_B = A m_n/V = A m_n/(4 \pi r_0^3 A/3) = 3 m_n^4 \alpha_z^3 / \pi
\end{equation}
if the nucleon number $A = B$, the nucleus radius $r = r_0 A^{1/3}$, and the mass
radius $r_0 = 1/2 m_n \alpha_z$ are used.

The QCD vacuum represented by massive gluons is quantized by the
maximum wavevector mode $N_R \approx 10^{30}$, the total gluon
number $N_G = 4 \pi N_R^3/3 \approx 10^{91}$, and the gluon number
density $n_G = \Lambda_{QCD}^3 \approx 10^{-2} \ \textup{GeV}^3
\approx 10^{39} \ \textup{cm}^{-3}$ \cite{Roh3}. Massless photons
are quantized by the maximum wavevector mode $N_\gamma \approx
10^{29}$ and the total photon number $N_{t \gamma} = 4 \pi
N_\gamma^3/3 \approx 10^{88}$. Massless phonons in the matter
space are quantized by the maximum wavevector mode (Debye mode)
$N_D \approx 10^{25}$ and the total phonon number $N_{t p} = 4 \pi
N_D^3/3 \approx 10^{75}$. Vacuum, matter, photon, and phonon
energies are also thermodynamically quantized. Quantum states of
vacuum, matter, photons, and phonons have average occupation
numbers $f_b = 1 /(e^{(E - \mu)/T} - 1)$ for gauge bosons, $f_f =
1 /(e^{(E - \mu)/T} + 1)$ for baryons, $f_\gamma = 1 /(e^{E/T} -
1)$ for photons, and $f_p = 1 /(e^{E/T} - 1)$ for phonons under
the assumption of free particles in thermal equilibrium. The
quantization unit of energy due to a gauge boson in strong
interactions is $\Lambda_{QCD}/N_R \simeq 10^{-31}$ GeV.

\subsection{Nuclear Magnetic Dipole Moment}

The nuclear magnetic dipole moment shows good illustration for
colorspin as well as isospin as the intrinsic angular momentum in
addition to spin as the intrinsic angular momentum. The electric
dipole moment of a nucleon disappears to satisfy the parity of
nucleus but it is not zero even though it is extremely small as
reflected in the $\Theta$ parameter.

The magnetic dipole moment of a nucleon comes from two sources, the intrinsic dipole moments of constituent quarks and
the orbital motions of quarks.
The three quarks are symmetric in the spatial parts of their wave functions, which have relative motion between them
in the $l =0$ states, so that no contribution to the magnetic dipole moment comes from quark orbital motion.

The magnetic dipole moment of a nucleus is given by
\begin{math}
\mu_i = \frac{e}{2 m_p} l_i
\end{math}
where $l_i$ is the orbital momentum of the i-th proton and $m_p$ is the proton mass.
The magnetic dipole moment for orbital motion may be rewritten by
\begin{math}
\mu_i = g_l l_i \mu_N
\end{math}
where the Lande g-factor $g_l = 1$ for a proton and $g_l = 0$ for a neutron with the definition of nuclear magneton
$\mu_N = e/m_p = 3.15 \times 10^{-17} \ \textup{GeV}/T$.
Similarly, the contribution from the intrinsic spin of each nucleon may be expressed in the form
\begin{equation}
\mu_i = \frac{e}{2 m_p} s_i = g_{s} s_i \mu_N.
\end{equation}
Since $s = 1/2$, the Lande spin g-factor for a free nucleon is
$g_{s}^p = 2 \mu_p/mu_N = 5.59$ for a proton and $g_{s}^n = 2
\mu_n/\mu_N = -3.83$ for a neutron \cite{Wink}. The g-factors are
different with $g_{s} = 2$ for a pointlike electron and $g_{s} =
0$ for a pointlike neutral particle. Formerly, these differences
between theoretical and experimental values were ascribed to the
cloud of pion mesons that surround nucleons, with positive and
neutral pions around protons and with negative and neutral pions
around neutrons.

According to the quark model, the magnetic dipole moment of the
proton is given by $\mu_p = 4 \mu_u/3 - \mu_d/3$ in terms of the
magnetic dipole moments of u and d quarks. The magnetic dipole
moment of the neutron is given by $\mu_n = 4 \mu_d/3 - \mu_u/3$.
Under the assumption that the masses of u and d quarks involved
are equal and their ratio of magnetic dipole moments is $\mu_u = -
2 \mu_d$, the ratio between the magnetic dipole moments of the
proton and neutron is $\mu_n /\mu_p = - 2/3$, which is in good
agreement with the observed value of $-1.913/2.793 = - 0.685$.
However, the quark model does not give any clue for the absolute
values of the magnetic dipole moments.

The intrinsic principal number $n_m = (n_c, n_i, n_s)$ in intrinsic space quantization is, on the other hand, introduced
as the analogy to the principal number $n$ in extrinsic space quantization and the intrinsic angular momenta are analogous to the extrinsic angular momentum
so that the total angular momentum has the form of
\begin{equation}
\vec J = \vec L + \vec S + \vec I + \vec C .
\end{equation}
One of the explicit examples to illustrate the extension of the
total angular momentum might be the nuclear magnetic dipole
moment: the g-factors of the proton and neutron are
respectively $g_{s}^p = 5.59$ and $g_{s}^n = - 3.83$, which are
shifted from $2$ and $0$ expected for pointlike particles. It is
strongly suggested in order to resolve the problem of the nuclear
magnetic moment that contributions from colorspin and isospin
degrees of freedom are included to nucleons. The shifted values
for the proton and neutron, $3.59$ and $- 3.83$ are almost
identical and they mostly come from the combined contribution of
colorspin, isospin, and spin. The mass ratio of the proton and the
constituent quark, $m_p/m_q \sim 2.79$, also represents combined
colorspin, isospin, spin degrees of freedom. The excess
contribution $|g_{s}| \simeq 3.83$ common for $\mu_p$ and $\mu_n$
comes from the magnetic dipole moment due to the $SU(2)_N \times
U(1)_Z$ symmetry for color charges and the $SU(2)_L \times U(1)_Y$
symmetry for isospin charges while the contribution $g_{s} \simeq
1.76$ only for $\mu_p$ comes from the magnetic dipole moment due
to the $U(1)_f$ gauge symmetry. This interpretation is easily
justified if the coupling constant $g_f = \sqrt{c_f^f \alpha_s} =
\sqrt{\alpha_s/16} \simeq 1.7 \ e$ for the $U(1)_f$ gauge theory
and the coupling constant $g_b = \sqrt{c_f^b \alpha_s} =
\sqrt{\alpha_s/3} \simeq 3.8 \ e$ for the $SU(2)_N \times U(1)_Z$
gauge theory. The description above reflects the mixed
contribution of colorspin, isospin, and spin degrees of freedom
and the total angular momentum $\vec J = \vec L + \vec C + \vec I
+ \vec S$: the total nuclear magnetic moment $\mu_{tN} = (g_s s +
g_n c + g_w i + g_l l) \mu_N$ with the respective coupling constants
$g_s$,$g_w$, and $g_n$ for sin, isospin, and colorspin interactions.

\subsection{Deuteron}

A deuteron (${^2}\textup{H}$) is the weak bound state of a proton
and a neutron. The ground state of the deuteron becomes the state
of strong isospin singlet $i^s = 0$ and spin triplet $s = 1$. The
reason why no bound states of proton-proton and neutron-neutron
are observed is explained by the violation of discrete symmetries
in color pairing mechanism. In strong interactions, the bindings
are regarded as color electric dipoles with intrinsic odd parity,
which must be suppressed. The deuteron state can also be slightly
modified if colorspin is taken into account. The ground state of
the deuteron may be the mixed state of two states possible by the
symmetry argument: one is the dominant state with weak isospin
singlet $i = 0$, spin triplet $s = 1$, and color singlet $c =0$
and the other is the contaminating state with weak isospin singlet
$i = 0$, spin singlet $s = 0$, and color triplet $c =1$.

The binding energy $- 2.23$ MeV is relatively less than the average binding energy for the other nucleus.
This is interpreted as that the interaction of
the deuteron is governed by the nuclear interaction of the $U(1)_Z$ gauge theory.
The coupling constant $\alpha^{a}_z = - \frac{\alpha_s}{6} = - 0.08$ for an asymmetric $U(1)_Z$ gauge symmetry seems to produce the
deuteron binding energy by
$E_{DB} = - \frac{\alpha_z^{a 2} m_n}{2} \frac{3}{4} = - 2.2$ MeV, which is
close to experimental value $- 2.23$ MeV.
The factor $3/4$ might come from the mixing of two states or from the
difference in the principal quantum number between the ground state and the first excite state.

In terms of the nuclear magneton $\mu_N$, the magnetic dipole
moment of the deuteron may be written as a function of the orbital
angular momentum operator and the intrinsic angular momentum
operator of each nucleon: $\mu_d = (g_p s_p + g_n s_n + l_p)
\mu_N$ where $l_p$ is the angular momentum of the proton. For $l_p
= 0$, the expectation value of the magnetic dipole operator
reduces to a sum of the intrinsic dipole moment of a proton and a
neutron: $\mu_d = \mu_p + \mu_n = 0.8798 \mu_N$ which is different
with the observed value of $0.8574 \mu_N$ \cite{Kell}. The small
difference $- 0.022339 \mu_N$ is conventionally expressed by the
mixing of d wave to s wave; the observed value is consistent with
$96$ percent s wave and $4$ percent d wave contributions. However,
the small discrepancy may alternatively be ascribed by the color
anomalous magnetic moment, which may be estimated by $(g^d_s-2)/2
\approx \alpha_n/2 \pi \approx 0.02$ with the coupling constant
$\alpha_n = \alpha_s/3 \simeq 0.12$ if the analogy of the electric
anomalous magnetic moment $(g_s-2)/2 \simeq \alpha_e/2 \pi$ is
used. The anomalous magnetic moment might come from the mixing of
two intrinsic states.

\subsection{Nucleon-Nucleon Scattering}

The nucleon-nucleon (NN) scattering is one of excellent examples
whose cross sections can be evaluated by QND as an $SU(2)_N \times U(1)_Z$
gauge theory. The ground states of
the proton and neutron are considered as a colorspin doublet in
QND. Quantum numbers of NN systems shown in Table \ref{nnqu} are
applied to analyze the NN scattering. Furthermore, it is realized
that the NN scattering has analogous properties with weak decay
process. In the scattering of the proton-neutron (pn),
proton-proton (pp), or neutron-neutron (nn), each cross section is
almost identical without the dependence of electric charge since
strong interactions are much stronger than electromagnetic
interactions. The NN scattering is basically the combination of
two interactions, which are explained by $SU(2)_N$ gauge theory at
relatively high energies and by the $U(1)_Z$ gauge theory at
relatively low energies. This investigation has a complete analogy
with the beta decay of the neutron in weak interactions, $n
\rightarrow p^+ + e^- + \bar \nu$; the distinction of Fermi type
and Gamow-Teller type scattering is possible.

The invariant amplitude in strong interactions as an $SU(2)_N$ gauge theory
at relatively high energies is given by
\begin{equation}
{\cal M} = - \frac{c^n_f g_s^2}{4} \frac{1}{k^2 - M_G^2} J_c^\mu J_{c \mu}^{\dagger}
\end{equation}
where the gluon mass $M_G$ is inserted in the gluon propagator.
This provides the cross section for the pp, nn, or pn scattering
as a colorspin triplet expressed by
\begin{equation}
\label{nnsc1}
\sigma = \frac{4 G_R^2 T^2}{\pi} \frac{1}{1 + 4 T^2/M_G^2}
\end{equation}
in the center of mass energy $T$ since ${\cal M} \sim G_R J_c^\mu J^\dagger_{c \mu}$
in terms of the effective strong coupling constant $G_R = \sqrt{2} c^n_f g_s^2/8 M_G^2$.
At high energy $T >> M_G$, the cross section converges to
\begin{math}
\sigma = \frac{4 G_R^2 M_G^2}{\pi}
\end{math}
while at low energy $T < M_G$, the cross section becomes
\begin{math}
\sigma = 4 G_R^2 T^2/\pi .
\end{math}
Cross sections for the NN scattering in terms of the massive gluon
exchange show excellent agreement with measurement data
\cite{Galb,Perk}.  According to strong isospin invariance, three
types of scattering such as the nn, pp, and pn scattering with
strong isospin one (spin zero) exhibit almost the same cross
sections. The theoretical cross section of the NN scattering as an
$SU(2)_N$ gauge theory at high energies about from $2$ GeV to
$10^3$ GeV is saturated to the experimental one of about $40$ mb
\cite{Galb} and the cross section at low energies from $0.6$ GeV
to $2$ GeV is roughly proportional to $T^2$: $c_f = 1/4$,
$\alpha_s = 0.48$, $M_G \approx 300$ MeV, and $G_R \approx 10 \
\text{GeV}^{-2}$ are used in this evaluation. The symmetric
$SU(2)_N$ colorspin interaction for the isospin triplet and spin
singlet contribution is commonly involved in the above three types
of nucleon-nucleon interactions with the massive gluon exchange.

In addition to the $SU(2)_N$ gauge symmetry, there is another
process, which explicitly appears at relatively low energies. The
cross section in strong interactions as a $U(1)_Z$ gauge theory at
relatively low energies is nonrelativistically obtained using the
Yukawa potential $V(r) = \sqrt{c^z_f \alpha_s} e^{- M_G (r -
l_{QCD})}/r$ with the strong scale $l_{QCD} = 1/\Lambda_{QCD}
\simeq 10 \ \textup{GeV}^{-1}$:
\begin{equation}
\label{nnsc2}
\sigma = 4 \pi \frac{c^z_f \alpha_s m_n^2}{M_G^4} \frac{1}{1 + 4 m_n T/M_G^2}
\end{equation}
where $c^z_f = - 1/6$ (or $1/12$) is the asymmetric (symmetric)
color factor, $m_n$ is the nucleon mass $m_n \simeq 940$ MeV, $T$
is the incident particle energy, and the gauge boson mass $M_G
\approx 140$ MeV. The cross section at relatively long range ($r >
2$ fm) is dependent on angular momentum and is well explained by
the massive gauge boson $M_G \sim m_\pi \sim 140$ MeV; this is
confirmed by a pion exchange in effective models. The cross
section data in the region below the energy $0.3$ GeV or above the
range $1.5$ fm are obtained by using the above formula, which is
definitely dependent on angular momenta. The comparison between
theoretical and measurement data shows good agreement. The
calculated cross section data in the limit $T \rightarrow 0$ give
agreement with observed data for cross sections ($\sigma = 4 \pi
a^2$) $\sigma_{pp} \simeq \sigma_{nn} \simeq 35$ b and
$\sigma_{pn} \simeq 66$ b for spin singlet and $\sigma_{pn} \simeq
4$ b for spin triplet as shown in Table \ref{nnsl} \cite{Wils}.
The effective (running) coupling constant $\alpha_z = c_f^z
\alpha_s = \alpha_s/12$ becomes stronger at lower energies but is
still less than $1$ so that the higher order corrections are
perturbatively possible: $\alpha_z \simeq 0.04$ around $100$ MeV.
A comparison of cross sections at low energies may give aid to
clarify such nuclear interactions. The cross section difference in
strong isospin triplet is mainly due to the contribution of
colorspins. The value of $a_{pn}^s = - 23.7$ fm is noticeably
larger than $a_{pp}$ and $a_{nn}$ at $s = 0$ and $i =1$ channel:
this is explained conventionally by the exchange of the charged
pion and the exchange of the neutral pion. This may alternatively
be explained by the $SU(2)_N$ gauge theory: there is only the
color triplet $c = 1$ contribution for $a_{pp}$ and $a_{nn}$ but
there are both color triplet $c = 1$ and singlet $c = 0$
contributions for $a_{pn}$ if color degrees of freedom are adopted
in addition to isospin and spin degrees of freedom. Therefore, the
additional contribution for the scattering length ($\sim -6$ fm)
is due to one of the color singlet configuration as an isospin
triplet and spin triplet state.  The mixing of intrinsic and
extrinsic waves is imposed as seen in the magnetic dipole moment
of the deuteron.

The NN scattering such as the pp, nn, or pn scattering with strong
isospin one (spin zero) is the Fermi type scattering in strong
interactions while the NN scattering such as the pn scattering or
np scattering with strong isospin zero (spin one) is the
Gamow-Teller type scattering in strong interactions; the concept
of the Fermi type scattering and Gamow-Teller type scattering is
just named after the concept of the Fermi decay and Gamow-Teller
decay in weak interactions. For the pn scattering, the
Gamow-Teller type scattering for an isospin singlet and spin
triplet state is expected in addition to the Fermi type scattering
as an isospin triplet and spin singlet state as seen in the
deuteron as an isospin singlet and spin triplet state. This is
explicitly shown in the observation of cross sections for the pn
scattering: the scattering length $a_{pn}^s \simeq - 23.7$ fm for
spin singlet as the Fermi type scattering and the scattering
length $a_{pn}^t = 5.4$ fm for spin triplet as the Gamow-Teller
type scattering.

There exist the color charged current mediated
by charged massive gluons $A^{\pm}$ with the mass $M_A$ and the color neutral current mediated by
neutral massive gluon $B^0$ with the mass $M_B$.
The NN scattering is good example to illustrate the existence of these massive gluons.
The invariant amplitude of the charged current is given by
\begin{equation}
\label{stcc} {\cal M}^{c} = - \frac{4 G_R}{\sqrt{2}} J_c^\mu J_{c
\mu}^{\dagger}
\end{equation}
and the invariant amplitude of the neutral current is given by
\begin{equation}
\label{stnc} {\cal M}^{n} = - \frac{4 G_R}{\sqrt{2}} 2 \rho
J_c^\mu J_{c \mu}^{\dagger} .
\end{equation}
It is identified that the relative strength of the color neutral and
charged current in strong interactions becomes
\begin{equation}
\rho = \frac{M_B^{2}}{M_A^{2} \cos^2 \theta_R} .
\end{equation}
The NN scattering data approximately show $\rho \simeq 1$ as confirmed by isospin
invariance: $\sigma_{pn} \simeq \sigma_{pp} \simeq \sigma_{nn}$ for isospin singlet
and $\cos ^2 \theta_R = \frac{M_A^{2}}{M_B^{2}}$.  Summarizing, the effective coupling
constant $G_R \simeq 10 \ \textup{GeV}^{-2}$ and the gauge boson mass $M_G \simeq 300$
MeV for the $SU(2)_N$ gauge theory can be evaluated from the saturated cross section
$\sigma = 40$ mb at high energy and the effective coupling constant $G_R \simeq 10 \
\textup{GeV}^{-2}$ and the gauge boson mass $M_G \simeq 140$ MeV for the $U(1)_Z$
gauge theory can also be evaluated from the lower cross section $\sigma = 23$ mb. The
cross section $\sigma \approx 35$ b as spin singlet or $\sigma \approx 4$ b as spin
singlet at low energy $(T \rightarrow 0$) is dependent on the lowest extrinsic angular
momentum ($l=0$). This implies that gluons with the higher mass play the dominant role
depending on the colorspin angular momentum at the short range ($r < 1$ fm) while
gluons with the lower mass play the dominant role depending on the orbital angular
momentum at the long range ($r > 1$ fm). It is thus emphasized that QND as the
$SU(2)_N \times U(1)_Z$ gauge theory produces the cross section data, which do not
have divergence problems so that QND is renormalizable, from the zero energy limit to
almost $10^{3}$ GeV.

\begin{figure}[htb]
\vspace{10pt} \centerline{ \vbox to 250pt{\vss
   \hbox to 500pt{\includegraphics{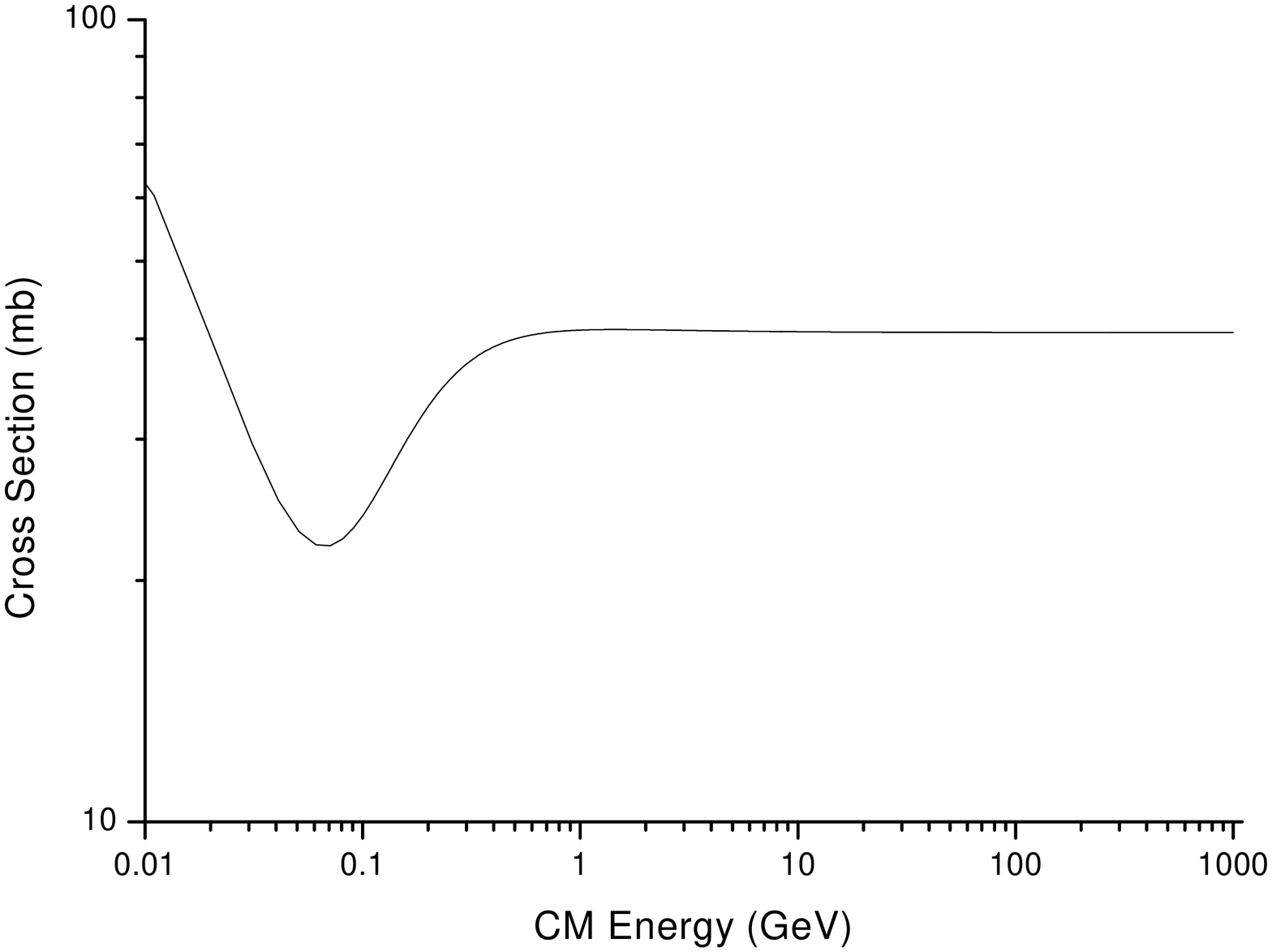}\hss}}}
\caption{The cross section of proton-proton scattering as a
function of the center of mass energy} \label{nnsc}
\end{figure}

Figure \ref{nnsc} shows the theoretical cross section of proton-proton scattering as a
function of the center of mass energy in terms of equations (\ref{nnsc1}) and
(\ref{nnsc2}): for example, the figure is obtained by using input data such as
the effective coupling constant $G_R = 71 \ \textup{GeV}^{-2}$ and gluon mass $M_{G1} =
0.27$ GeV for equation (\ref{nnsc1}) and such as the gluon mass $M_{G2} = 0.15$ GeV and
running coupling constant $\alpha_z = 0.02$ at $1$ GeV for equation (\ref{nnsc2}).
This illustrates good agreement in general trend and in absolute magnitude with the
experimental data, which are well shown as a function of the laboratory energy in
Perkins \cite{Perk}, in the energy scale from almost zero to $10^3$ GeV.  It is also
emphasized that only the coupling constant is, in principle, necessary as an input
parameter in the calculation of cross section. More fine tuning is required to
determine parameters such as the strong coupling constant $\alpha_s$ or $\alpha_z$,
gluon mass $M_G$, and effective coupling constant $G_R$ since the cross section data are
sensitive to input parameters.

\subsection{Meson-Nucleon Scattering}

QND as the $SU(2)_N \times U(1)_Z$ gauge theory may be applied to
the meson-nucleon scattering. At higher energies above $5$ GeV,
the cross section becomes the same for particle and antiparticle
according to the Pomerancuk theorem and it is moreover isospin
independent. Putting all these together, an expected cross section
ratio $\sigma (\pi N)/\sigma (NN) \simeq 2/3$ follows from simple
quark counting. The ratio $2/3$ might stem from the color factor
ratio of the meson to the baryon since the color factor of color
symmetric octet for the quark-antiquark interaction is $c_f = 1/6$
and the color factor of color symmetric sextet for the quark-quark
interaction is $c_f = 1/4$. The cross sections for the $\pi^- p$,
$\pi^+ p$, $\pi^- n$, and $\pi^- n$ scattering become almost equal
from charge independence: the measurement data show about $\sigma
\simeq 25$ mb. Likewise, the same argument is applied to cross
sections for the $K^- p$, $K^+ p$, $K^- n$, and $K^- n$
scattering: the measurement data show about $\sigma \simeq 20$ mb.
The experiment data for the meson-nucleon scattering as well as
the NN scattering illustrate massive gluons with about the mass
$M_G \simeq 300$ MeV responsible for QND. At lower energies, it is
conventionally known that cross sections for the $\pi^- p$ and
$\pi^+ p$ scattering are dominated by the $P_{33}$ channel ($l=1,
i=3/2, s=3/2$). This is not contradictory to the concept of colorspin
since the dominant channel may become the channel with quantum
numbers $c=1$, $i=3/2$, and $s=3/2$ by replacing the extrinsic
angular momentum $l=1$ with the intrinsic colorspin angular
momentum $c=1$ at relatively higher energies. This interpretation
may be possible since the lifetime of the resonance $\Delta^{++}$
particle shows the typical strong lifetime $\tau \simeq 1/G_R^2
m_\Delta^5 \simeq 10^{-23}$ sec and the cross section data except
resonance energies have the typical cross sections for strong
interactions $\sigma \simeq G_R^2 T^2 \simeq 30$ mb, originated by
massive gluons with about the mass $M_G \simeq 300$ MeV.

Other meson-nucleon interactions $\pi^\pm + p \rightarrow K^+ + \Sigma^\pm$,
$\pi^- + p \rightarrow K^0 + \Sigma^0$, $\pi^- + p \rightarrow K^0 + \Lambda$,
$K^- + p \rightarrow \pi^0 + \Lambda$, etc.
similarly indicate strong interactions by the exchange of massive gluons.
In these interactions, the reaction $d \bar d \rightarrow s \bar s$
is commonly included in the quark level.
They show the typical cross sections  for strong interactions and
conservation law for the strangeness number as noted by electric charge
quantization $Q_f = C_3 + Z_c/2$ with $Z_c = B + S$.
The strong decay $\Sigma^\pm (1385) \rightarrow \Lambda + \pi^\pm$ holding
the typical lifetime and decay width is also explained by the exchange of
massive gluons.

If the description above is correct, the meson-nucleon scattering
as well as the hadron decay can be studied over much wider energy
range in terms of the unified view of QND.

\subsection{Nuclear Potential}

The nuclear potential is discussed from QND as a gauge theory point of view;
the derivation and extension of each nuclear potential term are in principle possible.
The colorspin-colorspin interaction is introduced as the central potential and
the isospin-orbit potential and colorspin-orbit potential are introduced as non-local potentials
in analogy with the spin-orbit potential.
Colorspin and isospin as internal degrees of freedom play roles of intrinsic angular momenta.

The nuclear central potential is generally expressed by
\begin{equation}
\label{nuce}
V_c (r) = V_0 + V_S \vec \sigma^i \cdot \vec \sigma^j +
V_I \vec \tau^i \cdot \vec \tau^j + V_{S I} \vec \sigma^i \cdot \vec \sigma^j \vec \tau^i \cdot \vec \tau^j
\end{equation}
where the first term denotes the pure radial distance dependent
potential, the second term the spin-spin interaction, the third
the strong isospin-isospin interaction, and the fourth the
spin-spin and isospin-isospin interaction. This can be generalized
as follows if strong isospin is decomposed with colorspin and weak
isospin:
\begin{eqnarray}
\label{nuce1} V_c (r) & = & V_0 + V_C \vec \zeta^i \cdot \vec \zeta^j + V_S \vec
\sigma^i \cdot \vec \sigma^j + V_I \vec \tau^i \cdot \vec \tau^j \nonumber \\ & + &
V_{CI} \vec \zeta^i \cdot \vec \zeta^j \vec \tau^i \cdot \vec \tau^j  +  V_{CS} \vec
\zeta^i \cdot \vec \zeta^j \vec \sigma^i \cdot \vec \sigma^j + V_{SI} \vec \sigma^i
\cdot \vec \sigma^j \vec \tau^i \cdot \vec \tau^j  \nonumber \\ & + & V_{CSI} \vec
\zeta^i \cdot \vec \zeta^j \vec \sigma^i \cdot \vec \sigma^j \vec \tau^i \cdot \vec
\tau^j
\end{eqnarray}
where fine interactions are colorspin-colorspin, (weak)
isospin-isospin, spin-spin interactions and hyperfine interactions
are colorspin-colorspin and isospin-isospin, colorspin-colorspin
and spin-spin, spin-spin and isospin-isospin interactions, and the
combination interaction of colorspin, isospin, and spin. Fine and
hyperfine interactions appear in several places; the nuclear
magnetic moment, the nucleon-nucleon scattering, and the nucleus
decays of $\Delta$ and $\Sigma^0$ are a few examples as discussed
in the previous subsections. This is consistent with the concept
of the intrinsic angular momentum $\vec J = \vec C + \vec I + \vec
S$ and central potentials between intrinsic angular momenta.
The potential $V_0$ is relevant for the $U(1)_Z$ gauge theory,
the colorspin-colorspin interaction for the $SU(2)_N$ gauge theory,
the isospin-isospin interaction for the $SU(2)_N$ gauge theory,
and the spin-spin potential for the $SU(2)$ spin symmetry.

The nuclear non-local spin-orbit potential may be expressed by
\begin{equation}
\label{nuls}
V_{LS} = V_{LS} (r) \vec L \cdot \vec \sigma
\end{equation}
which might have the form of
$V_{LS} \propto \alpha_f |\psi (0)|^2/m_i m_j$.
The spin-orbit interaction plays an important role in the formation of nucleus as shown in the shell model.
The nuclear spin-orbit coupling splits a nuclear energy into two levels $j = l \pm s$ except for $l = 0$,
but $j = l - 1/2$ lies higher in energy and $j = l + 1/2$ lies lower in energy.
The order of the levels is inverted from that in an atom because of the change of sign of the spin-orbit coupling in nuclei
compared to atoms.
This is the explicit evidence for the difference between angular momenta of atom and nucleus.
The nuclear non-local isospin-orbit potential may be analogously added if isospin is the intrinsic angular momentum like spin:
\begin{equation}
\label{nuls1}
V_{LI} = V_{LI} (r) \vec L \cdot \vec \tau
\end{equation}
which might have the form of
$V_{LI} \propto \alpha_f |\psi (0)|^2/m_i m_j$.
The isospin-orbit interaction plays a major role in the level splitting
between protons and neutrons and the level splitting among isobaric nuclei as shown in the shell model.
The nuclear isospin-orbit coupling splits a nuclear energy into two levels $j = l \pm i$,
but the neutron with $j = l - 1/2$ lies lower in energy and the proton with $j = l + 1/2$ lies  higher in energy as expected in the shell model.
The isospin-orbit coupling is also relevant to the splitting of the energy levels for isobaric nuclei.
The nuclear non-local colorspin-orbit potential may be analogously added if colorspin is also regarded as the intrinsic angular momentum like spin:
\begin{equation}
\label{nuls2}
V_{LC} = V_{LC} (r) \vec L \cdot \vec \zeta
\end{equation}
which might have the form of
$V_{LC} \propto \alpha_f |\psi (0)|^2/m_i m_j$.
The colorspin-orbit interaction in nucleus is significant at the surface of the nucleus.
For example, it prevents the harmonic oscillator potential from diverging at long range in the shell model.

\subsection{Shell Model}

The nuclear shell model \cite{Maye} is to describe nucleon states in terms of nucleon degrees of freedom
and the existence of magic numbers may be explained by independent particle model.
The $U(1)_Z$ gauge theory, apart from electromagnetic interactions due to the $U(1)_f$ gauge theory, may be very useful in studying the scattering, decay, and excitation of nuclei.
It provides quantum numbers such as the radial principal number $n$, the angular momentum number $l$, and the third component number of the angular momentum $m_{l}$ but
they have different origin compared to quantum numbers, ($n_e, l_e, m_{le}$), associated with quantum electromagnetism (QED) as a $U(1)_e$ gauge theory.
The average potential of nuclear matter is the sum of Yukawa potentials for individual particles and it creates
quantum numbers $(n, l, m_{l})$.
This is supported by the fact that the
orbital angular momentum $l$ of the nucleon has the different origin from the color
charge with the orbital angular momentum $l_e$ of the electron from the isospin charge
since two angular momenta have opposite directions from the information of spin-orbit
couplings in nucleus and atoms.

The independent particle shell model is roughly successful in the
magnetic dipole moment. Including both $l$ and $s$ terms, the
magnetic dipole moment becomes
\begin{math}
\label{dipo0}
\mu = (g_l l_z + g_{s} s_z) \mu_N
\end{math}
with $j = j_z$.
Taking the expectation value, the result is
\begin{equation}
\label{mgmo}
\langle \mu \rangle = [ g_l j + (g_{s} - g_l) \langle s_z \rangle] \mu_N
\end{equation}
since the expectation value of $s_z$ is given by $\langle s_z \rangle =
\frac{j}{2 j (j+1)} [j(j+1) - l(l+1) + s(s+1)]$: for $j = l +
1/2$, $\langle s_z \rangle = 1/2$, while for $j = l - 1/2$, $\langle s_z \rangle = -j/2(j+1)$.
The corresponding magnetic moments are expressed by $\langle \mu \rangle = [ g_l
(j - 1/2) +  g_{s}/2] \mu_N$ for $j = l + 1/2$ and $\langle \mu \rangle = [ g_l
\frac{j(j+ 3/2)}{j + 1} - g_{s} \frac{1}{2(j+1)}] \mu_N$ for $j =
l - 1/2$. Theoretical values as shown in Schmidt lines
($\langle \mu \rangle/\mu_N$) for odd-A nuclei show the success in the general
trend of observed values but experimental values are overall
agreement for $g_{s} = 0.6 g^f_{s}$ with the g-factor of the free
nucleon $g_{s}^f$. The factor $0.6$ might indicate the
contribution of color degrees of freedom to the magnetic dipole
moment since the actual spin g-factor $g_{s}$ in (\ref{mgmo}) may
be reduced due to the contribution from colorspin angular momentum
to the total angular momentum just as shown in its reduction in the deuteron.

In prospect, it may be possible to establish a periodic table for nuclei
just like the periodic table for atoms in terms of nuclear extrinsic quantum numbers $(n, l, m_{l})$ and magic numbers
if colorspin degrees of freedom in nucleons are taken into account.

\subsection{Nucleus Binding Energy}

QND may be applied to interpret the semi-empirical mass formula and eventually extended to evaluate
nucleus mass spectra as well as excitation energies.
In this part, only the binding energy for nucleus is briefly discussed how to apply QND to nucleus.

Nucleus mass formula is expressed by
\begin{equation}
m_N = Z \ m_p + N \ m_n + E_B
\end{equation}
where $Z$ is the proton number, $N$ is the neutron number, and $E_B$ is the total binding energy.
The form of this mass formula is very analogous to the mass formula for the hadron by the constituent quark model
as discussed in the previous section.
The binding energy $E_B$ semi-empirically constitutes of the volume, surface, Coulomb, symmetric, and pairing terms \cite{Beth}:
\begin{equation}
E_B (Z, N) = a_v A + a_s A^{2/3} + a_c Z^2/A^{1/3} + a_i (N - Z)^2/A - \delta (A)
\end{equation}
where $a_v \simeq - 16$ MeV, $a_s \simeq 18$ MeV, $a_c \simeq 0.7$
MeV, $a_i \simeq 28$ MeV, and the pairing term $\delta$. The
binding energy $E_B$ may be rewritten in terms of the extrinsic
principal number $n$ by
\begin{equation}
E_B (n) = a_v n^6/3 + a_s n^4 + a_c Z^2/n^2 + a_i (N - Z)^2/n^6 - \delta (n)
\end{equation}
since $A = n^6$.

The color $SU(3)_C$ symmetry generates the $SU(2)_N \times U(1)_Z$ symmetry or the
$U(1)_f$ symmetry, which governs nuclear dynamics. In terms of $\alpha_s \simeq 0.48$, the
nuclear coupling constant of the nucleon-nucleon interaction becomes $- f_n^2 =
- G_R m_n^2 = - \sqrt{2} c_f g_s^2 m_n^2/ 8 M_G^2 \approx - 15.5$ with the nucleon
mass $m_n = 0.94$ GeV and $G_R = \sqrt{2} c_f g_s^2/8 M_G^2$. The asymmetric
colorspin-colorspin interaction between nucleons might be the major contribution of
the volume term representing the saturation: $V_c = \sum_{i >j} V_c (r) \vec \zeta^i
\cdot \vec \zeta^j$, which is proportional to the limited number of nucleons $A$ and
the energy per particle approximately becomes constant. This reflects the fundamental
symmetry of nucleon force, which might be related to the saturation of the cross
section between nucleons at higher energy than the gluon mass. The color neutral
coupling constant $\alpha_b = \alpha_n/\cos \theta_R^2 = - \frac{2}{3} \alpha_s = -
0.32$ produces the depth of nuclear effective potential in terms of the hydrogen atom
analogy, $V_0 = - \frac{\alpha_b^2 m_n}{2} = - 48$ MeV, which might be relevant for
the depth of the Woods-Saxon potential $V_0 \approx - 50$ MeV: this can be also produced by
the constituent quark mass and g-factor, that is, $V_0 = - g \frac{\alpha_b^2 m_u}{2}
\approx - 48$ MeV with the mass ratio $g = m_n/m_u  \approx 2.79$. The binding energy
of the nucleon in infinite nuclear matter is obtained by $E_{B} = - \frac{\alpha_b^2
m_u}{2} = - 18.3$ MeV in terms of the constituent quark mass $m_u$: it corresponds to
the binding energy due to the volume contribution in the semi-classical mass formula.
The surface energy might originate from the colorspin-orbit coupling, in which the cross
section between nucleons increases as the energy increases. The symmetry energy is
roughly understood when Fermi-gas model, where nucleons are treated like
non-interacting particles, are used but it can be studied as color interactions as a
color doublet of the proton and neutron.

The description above may be easily estimated as follows. The
empirical binding energy $- 2.23$ MeV for the deuteron is governed
by the nuclear interaction as the $U(1)_Z$ gauge theory. The
average binding energy for the nuclei greater than the nuclear
mass number $A \approx 12$ may be obtained by $-2.23 \times 4 =
-8.9$ MeV since there are four gauge bosons if their binding energy
is governed by the nuclear interaction as the $SU(2)_N \times
U(1)_Z$ gauge theory. The coefficient of binding energy due to
infinite nuclear volume is also estimated by $-2.23 \times 8 =
-17.8$ MeV since there are eight gauge bosons if their binding
energy is governed by the nuclear interaction as the $SU(3)_C$
gauge theory; the volume binding energy is proportional to the
nuclear mass $A$. The coefficient of binding energy due to nuclear
surface is also estimated by $8.9 \times 2 = 17.8$ MeV since there
are two degenerated fermions if their surface binding energy is
governed by the intrinsic and extrinsic orbit interactions: the
surface energy is proportional to the average of the potential
$V_{LC} \vec L \cdot \vec \zeta $. The depth of the average
potential is obtained by $-17.8 \times 2.79 = -49.6$ MeV with the
mass ratio $g = 2.79$. The estimated values are reasonably closed
to observed values.

\section{Comparison between Quantum Nucleardynamics and Effective Models}

In the previous section, nuclear phenomena supporting QND are addressed and
scientific merits of QND compared with nuclear effective models are presented in this section.
This section is thus devoted to summarize and to convince QND
applicable to whole nuclear phenomena.

Comparison between QND and
effective models is given in Table \ref{comp}. Effective models,
regardless of nonrelativistic or relativistic, stand for Bonn,
Paris, Reid phenomenological potentials, meson exchange models,
Walecka model \cite{Wale}, linear sigma model \cite{Gell3},
nonlinear sigma model \cite{Wein5}, Nambu-Jona-Lasinio model
\cite{Namb3}, chiral effective models \cite{Gass}, bag model
\cite{Chod}, quark model, QCD sum rule \cite{Shif}, Skyrme model
\cite{Skyr}, constituent quark model \cite{Dash}, other QCD
inspired models, etc.

QND as an $SU(2)_N$ gauge theory has the
similarity to the effective models with the omega meson ($\omega$)
exchange and QND as a $U(1)_Z$ gauge theory has the similarity to
the effective models with the pion exchange. Short range repulsion
in the nucleon-nucleon interaction ($r < 0.5$ fm) is due to the
gluon $A_3$ in the symmetric configuration of quark-quark
interactions and long range attraction ($r \sim 1$ fm) is due to
the gluon $A_8$ in the asymmetric configuration of
quark-quark interactions; this is related to the relativistic
effective models since the roles of $A_8$ are similar to the
attractive roles of the pion meson ($\pi$) and sigma meson
($\sigma$), respectively, and the role of $A_3$ is similar to the
repulsive role of the omega meson ($\omega$) \cite{Wale}. The
coupling constant for repulsion is predicted by $\alpha_z =
\alpha_s/12 \approx 0.04$ at the QCD cutoff. This scheme provides
the derivation of nuclear parameters, which can be used to study
the nucleon-nucleon interaction, and the periodic table of nuclear
properties, which are characterized by the shell model, from the
viewpoint of the microscopic theory for strong interactions, QCD.

QND is applicable to various aspects in the wide energy range but effective
models are only effective to few aspects of nuclear phenomena in the rather small
energy range.  The effective strong coupling constant $G_R/\sqrt{2}= c_f
g_s^2/8 M_G^2 \approx 10 \ \textup{GeV}^{-2}$ like Fermi weak constant $G_F/\sqrt{2} =
g_w^2/8 M_W^2 \approx10^{-5} \ \textup{GeV}^{-2}$ are used to study nuclear
interactions. The proton number conservation is the result of the $U(1)_f$ gauge
theory and the baryon number conservation ($B = N_B \approx 10^{78}$) is the
consequence of the $U(1)_Z$ gauge theory for strong interactions: the charge
quantization $\hat Q_f = \hat C_3 + \hat Z_c/2$ with the hyper-color operator $\hat
Z_c = \hat B + \hat S$ for the $U(1)_f$ gauge theory. The mass density and charge
density of nuclear matter are the consequence of the proton number conservation or the
baryon number conservation. In analogy with the electron orbit radius of the hydrogen
atom $r_e = a_0 n_e^2$ with the atomic radius $a_0 = 1/2 m_e \alpha_y$ and the
principal quantum number $n_e$, the nuclear radius may be written by $r = r_0 n^2$
with a nucleon orbit radius $r_0 = 1/2 m_n \alpha_z \approx 1.2$ fm and the principal
quantum number $n = A^{1/6} = B^{1/6}$. The proton and neutron are assigned as a
colorspin plus weak isospin doublet instead of as a strong isospin doublet. The
magnetic dipole moment and electric quadrupole moment further suggest that colorspin
and isospin may be introduced as intrinsic angular momenta in addition to the spin
angular momentum. The extension of the total angular momentum $\vec J = \vec L + \vec
S + \vec C + \vec I$ from  $\vec J = \vec L + \vec S$ may explain the Lande spin
g-factors of magnetic dipole moments for the proton and neutron, $g_s^p = 5.59$ and
$g_s^n = - 3.83$, respectively. The nucleon-nucleon scattering can be investigated in
terms of QND as an $SU(2)_N \times U(1)_Z$ gauge theory in analogy with the beta decay
of the neutron in weak interactions. The nucleon-nucleon scattering (pp, nn, or np) as
a spin singlet is known as the Fermi type scattering and the nucleon-nucleon
scattering (np or pn) as a spin triplet is known as the Gamow-Teller type scattering.
Further testable predictions or already confirmed predictions from QND are given as
follows: parity violation in meson and baryon spectra, charge conjugation violation in
baryon spectra, time reversal and CP violation in the electric dipole moment for the
neutron ($\Theta \leq 10^{-9}$), the nonconservation of color singet proton and
neutron, the nonconservation of the axial vector current, coupling constant hierarchy,
the color mixing angle $\sin^2 \theta_R \simeq 1/4$, the assignment of strong isospin
as colorspin plus weak isospin. Moreover, QND possessing colorspin degrees of freedom
may apply to explain the various nuclear phenomena such as lifetimes and cross
sections of nuclear scattering and reaction, shell model, meson-nucleon scattering,
nuclear potential, nuclear binding energy, gamma ray, etc. over the wide energy range.

\section{Conclusions}

This paper claims that QCD as an $SU(3)_C$ gauge theory produces
quantum nucleardynamics (QND) as an $SU(2)_N \times U(1)_Z$ gauge theory
in terms of dynamical spontaneous symmetry breaking
(DSSB) through the condensation of singlet gluons.
Important concepts are colorspin, massive gauge boson, and discrete
symmetry breaking.
The proton and neutron assigned as a strong isospin doublet are
identified as a colorspin plus weak isospin doublet; nucleons as
color doublets are governed by QND.
Quantized gauge bosons, i.e., gluons, are massive and yield the Yukawa
potential; this is understood as the confinement mechanism of massive gluons limited
to relatively short range propagation. Phase transition occurs when singlet gluons,
rather than other scalar bosons, acquire vacuum expectation values and reduce the
masses of gauge bosons: the gauge boson mass square is $M^2_G = M^2_{H} - c_f g_s^2
\langle \phi \rangle^2 = c_f g_s^2 [A_{0}^2 - \langle \phi \rangle^2]$ with the grand unification mass $M_H$, the
strong coupling constant $g_s$, and the singlet gluon condensation $\langle \phi \rangle$.

QND as an $SU(2)_N \times U(1)_Z$ gauge theory or as a $U(1)_f$ gauge theory is
applied to evaluate lifetimes and cross sections in strong interactions or to study
alpha and gamma decays. QND is applicable to study the nucleon-nucleon scattering in
analogy with the beta decay of the neutron in weak interactions: the Fermi type
scattering and Gamow-Teller type scattering in strong interactions are suggested just
like the Fermi decay and Gamow-Teller decay in weak interactions. It is also useful to
the meson-nucleon scattering so that cross sections over wider energy regions are
analyzed by the exchange of massive gluons. The strong isospin symmetry and magnetic
dipole moment for the proton and neutron suggest the introduction of the colorspin
intrinsic angular momentum. The color factors described above are the color factors
purely due to color charges but the effective color factors used in nuclear dynamics
must be multiplied by the isospin factor $i_f^w = \sin^2 \theta_W = 1/4$ since the
proton and neutron are a colorspin and isospin doublet: $c_f^{eff} = i_f^w c_f = i_f^w
(c_f^b, c_f^n, c_f^z, c_f^f) = (1/12, 1/16, 1/48, 1/64)$ for symmetric configurations.
The conservation of the proton number is the result of the $U(1)_f$ local gauge theory
just as the conservation of the electron number is the result of the $U(1)_e$ local
gauge theory. The baryon number conservation is the result of the $U(1)_Z$ gauge
theory for strong interactions just as the lepton number conservation is the result of
the $U(1)_Y$ gauge theory for weak interactions. The constant mass density and charge
density of nuclear matter are the consequence of the proton number conservation or the
baryon number conservation. The colorspin-colorspin interaction potential is
introduced in the nuclear local central potential as the major contribution and the
colorspin-orbit coupling potential and isospin-orbit coupling potential as non-local
potentials are suggested in addition to the spin-orbit coupling potential. Several
contributions of nuclear binding energies are also interpreted from the gauge theory
point of view. This proposal thus provides the connection between QCD and QND, which
are both renormalizable just like the GWS model. This implies that nuclear dynamics
can be studied from the gauge theory without free parameters except the strong
coupling constant. Lots of applications for QND are expected toward the complete
understanding of nuclear matter.

Notable consequences of this work are summarized in the following. QCD as an $SU(3)_C$
gauge symmetry generates QND as an $SU(2)_N \times U(1)_Z$ gauge theory or a $U(1)_f$
gauge theory through DSSB induced by the condensation of singlet gluons. The
nonperturbative solution of QCD in the low energy limit is explicitly obtained from
DSSB. The concepts of colorspin, massive gluon, and discrete symmetry breaking play
important roles in nuclear interactions. The analogy property emphasizes that QND is
the analogous partner of the GWS model and that massless gauge bosons mediate nuclear
electromagnetic interactions. QND as an $SU(2)_N \times U(1)_Z$ gauge theory emerges
as the dynamics for nuclear interactions so that cross sections, lifetimes,
nucleon-nucleon scattering, meson-nucleon scattering, magnetic dipole moment, nuclear
potential, nuclear binding energy, gamma decay, etc are studied. The baryon number
conservation is the consequence of the $U(1)_Z$ gauge theory and the proton number
conservation is the consequence of the $U(1)_f$ gauge theory.  This proposal may also
provide a turning point toward the understanding of nuclear forces at relatively low
energies since QND is more or less applicable to all the nuclear phenomena.

\onecolumn

\newpage

\begin{table}
\caption{\label{coqu} Color Quantum Numbers of Nucleons}
\end{table}
\centerline{
\begin{tabular}{|c|c|c|c|c|} \hline
Baryons & $C$ & $C_3$ & $Z_c$ & $Q_f$ \\ \hline \hline
$p_d$ & $1/2$ & $1/2$ & $1$ & $1$ \\ \hline
$n_d$ & $1/2$ & $-1/2$ & $1$ & $0$ \\ \hline
$p_s$ & $0$ & $0$ & $2$ & $1$ \\ \hline
$n_s$ & $0$ & $0$ & $0$ & $0$ \\ \hline
\end{tabular}
}

\vspace{1cm}

\begin{table}
\caption{\label{coga} Relations between Conservation Laws and
Gauge Theories}
\end{table}
\centerline{
\begin{tabular}{|c|c|c|} \hline
Force & Conservation Law & Gauge Theory \\ \hline \hline Electromagnetic & Proton &
$U(1)_f$ \\ \hline Strong & Baryon & $U(1)_Z$ \\ \hline Strong & Color Vector &
$SU(2)_N \times U(1)_Z$ \\ \hline Strong & Color & $SU(3)_C$ \\ \hline Electromagnetic
& Electron & $U(1)_e$ \\ \hline Weak & Lepton & $U(1)_Y$ \\ \hline Weak & V-A &
$SU(2)_L \times U(1)_Y$ \\ \hline
\end{tabular}
}

\vspace{1cm}

\begin{table}
\caption{\label{nnqu} Quantum Numbers of Nucleon-Nucleon Systems
($i^s$: strong isospin, $i$: weak isospin)}
\end{table}
\centerline{
\begin{tabular}{|c|c|c|} \hline
State & $i^s=1$  & $i^s=0$ \\ \hline \hline
pp & $i=1, s=0, c=1$ &  \\ \hline
nn & $i=1, s=0, c=1$ &  \\ \hline
pn & $i=1, s=0, c=1$ & $i=0, s=1, c=0$ \\
   & $i=1, s=1, c=0$ & $i=0, s=0, c=1$ \\ \hline
\end{tabular}
}

\vspace{1cm}

\begin{table}
\caption{\label{nnsl} Nucleon-Nucleon Scattering Length $a$ (fm)
and Effective Range $r_e$ (fm)}
\end{table}
\centerline{
\begin{tabular}{|c|c|c|} \hline
Scattering & $i^s=1, s=0$ & $i^s=0, s=1$ \\ \hline \hline
pp & $a=-17.1$,$r_e=2.79$ &  \\ \hline
nn & $a=-16,6$,$r_e=2.84$ &  \\ \hline
pn & $a=-23.7$,$r_e=2.73$ & $a=5.4$, \ $r_e=1.73$ \\ \hline
\end{tabular}
}

\newpage

\begin{table}
\caption{\label{comp} Comparison between Quantum Nucleardynamics
and Effective Models}
\end{table}
\centerline{
\begin{tabular}{|c|c|c|} \hline
Classification & QND & Effective Models \\ \hline \hline Exchange Particles & massive
gluons & massive mesons (model dependent) \\ \hline DSSB & yes & no \\ \hline Discrete
symmetries (P, C, T, CP) & breaking &  no \\ \hline Confinement & yes & no \\ \hline
$\Theta$ vacuum & yes & no
\\ \hline Baryon number conservation & $U(1)_Z$ gauge theory ($N_B
\simeq 10^{78}$) & unknown \\ \hline Proton number conservation &
$U(1)_f$ gauge theory & unknown \\ \hline Nuclear electromagnetic
interaction & $U(1)_f$ gauge theory & no \\ \hline Intrinsic
angular momenta & colorspin, weak isospin, spin & spin
\\ \hline Hadron mass generation & yes & unknown \\ \hline NN
scattering cross section & $G_R^2 T^2$ (from high to low energy) &
only at low energy \\ \hline Hadron decay time & $1/G_R^2 m_h^5$ &
no \\ \hline Neutron electric dipole moment & $\Theta \simeq
10^{-12}$ & no \\ \hline Free parameters & coupling constant &
many \\ \hline Renormalization & yes & model dependent \\ \hline
\end{tabular}
}

\end{document}